\documentstyle[12pt,fleqn]{article}
\begin{document}
\baselineskip= 22 truept
\newcommand{\nc}{\newcommand}
\begin{flushright}
IMSc-96/09/27\\
hep-th/ 9609152\\
\end{flushright}
\vspace{1cm}\begin{center} 
{\large \bf S-Duality and Exact Type IIB 
Superstring Backgrounds}\\
\vspace{1cm} 
{\bf {\sc Gautam Sengupta}}\\
\vspace{0.5cm}
{\it Institute of Mathematical Sciences, \\
CIT Campus, Tharamani\\
Madras 600 113\\
INDIA.\\
e-mail: gautam@imsc.ernet.in\\}
\bigskip
\end{center} 
\thispagestyle{empty}
\vskip 2cm
\begin{abstract}

A geometrical approach in the non-symmetric connection 
framework is
employed to examine the issue of higher order $\alpha^{\prime}$
corrections to D=10 type IIB superstring backgrounds with a 
covariantly 
constant null Killing isometry and non-zero Ramond-Ramond field 
content. 
These describe generalized supersymmetric string waves and were 
obtained
recently by us through the S-duality transformations of 
purely NS-NS
plane wave backgrounds. We find that the backgrounds are exact 
subject to the
existence of certain field redefinitions and provided certain 
restrictive
conditions are satisfied.

\end{abstract}
\vfil
\eject

\nc {\ap}{$\alpha^{\prime}$}
\nc {\be}{\begin{equation}}
\nc {\ee}{\end{equation}}
\nc {\bea}{\begin{eqnarray}}
\nc {\eea}{\end{eqnarray}}
\nc {\gam} {\Gamma}
\nc {\hi}{H_{\lambda\mu\nu}}
\nc {\pa}{\partial}
\nc {\rc}{R_{\lambda\mu\nu\kappa}}
\nc {\rci}{R^{+}_{\lambda\mu\nu\kappa}}
\nc {\ric}{R_{\mu\nu}}
\nc {\hf}{{1\over 2}}
\nc {\pab}{{\bar \partial}}
\nc{\rup}{R^{\rho}_{~\sigma\mu\nu}}
\nc{\rplus}{R^{+\rho}_{~~\sigma\mu\nu}}

\noindent {\bf 1. Introduction}
\bigskip

\noindent The past two years have witnessed remarkable developments
in the understanding of non-perturbative aspects of 
superstring theories.
Although a clear conception of the dynamical issues in the 
non-perturbative
regime is yet to emerge, symmetry considerations have yielded an
exciting glimpse. The central role in these investigations have been 
played by the
{\it duality symmetries} of superstring theories, 
both the perturbative
T-dualities \cite {tdual} and the non-perturbative
S \cite {sdual,senrev,jhs} and U \cite {hull} dualities. 
The latter relates weak and strong coupling regimes and
electricaly charged perturbative
string states with magneticaly charged solitons.
They arise as global non-compact symmetries in the low
energy effective supergravity theories which are conjectured 
to extend to
the full quantum superstring theory as a discrete version, 
being broken by
instanton effects. As an example we have the global
$SL(2,R)$ symmetry of the type IIB supergravity in ten dimensions 
\cite {hull,berg} extending
to the full type IIB superstring theory as $SL(2,Z)$ 
in ten dimensions, 
and relating
the NS-NS sector to the R-R sector. This leads to the possibility of
generating backgrounds with R-R field contents starting from purely
NS-NS configurations. We will refer to both the classical continuous
symmetry and the discrete symmetry of the quantum string theory 
as S-duality.

The R-R field content of the type II superstrings were not 
amenable to
a sigma model analysis \cite {sigmarev} as they lacked a conformal
field theory description on the world sheet \cite {fms}.
Very recently the conformal
field theory for the superstring backgrounds which are charged 
under the
R-R gauge fields have been identified to be that of open superstrings
with Dirichlet boundary conditions \cite {pol}. However such an 
interpretation for backgrounds uncharged under the 
R-R sector for {\it eg.} 
supersymmetric plane waves is still obscure.
The tree level equations of motion for these 
type II backgrounds, are of course
those of the type II supergravity which are field theory 
limits of type II
superstrings. In contrast, the higher order contributions to the 
tree level equations of
motion are still intractable for these non-p-brane like 
backgrounds, owing 
to the lack of an explicit sigma model description.
For recent results see refs. \cite {dim}.
Certain classes of string backgrounds
are found to have all higher order contributions to the 
equations of motion 
to be identicaly zero \cite {tset1,berg1,tsetk,tset2,tsetrev}
. Consequently the tree level equations are exact.
These higher order arguments are based on sigma model 
considerations and
are hence not applicable to the R-R sector of the non-brane like
backgrounds considered here.
However, there are a class of
string backgrounds with a covariantly constant null Killing isometry 
\cite {tset1,tsetk}
for the bosonic and heterotic versions of
which, there exists a purely geometrical approach to demonstrate that
the higher order terms are vanishing \cite {berg1,horo,duval,guv}.
This analysis makes no reference
to sigma models and are thus particularly suitable for the R-R sector
of type IIB superstring backgrounds.
The simplest
example in this class are the plane waves and the general class 
is referred
to as the K-models \cite {tsetk}. The K-model backgrounds posess 
vector like 
couplings
on the world sheet which on compactification provides background gauge
fields. The bosonic and heterotic K-models describe strings propagating
in a uniform magnetic field \cite {russ} which may be formulated as an
exact conformal field theory.

In an earlier article \cite {asg}, we had demonstrated that 
starting from a simple
plane wave background \cite {horo} with a purely NS-NS configuration
it is possible to generate non trivial R-R field content 
using the $SL(2,R)$
S-duality symmetry of the type IIB supergravity equations of 
motion in ten dimensions. This results in type IIB plane wave
backgrounds.
Furthermore using the geometrical considerations referred earlier, 
it could be 
shown that these type IIB backgrounds were exact
to all orders in \ap. 
This analysis was 
subsequently generalized \cite {ag} to the case of the 
K-models \cite {tsetk,tsetrev} ( with a Brinkmann
metric \cite {brink})  
and a flat transverse part in ten dimensions. We showed 
using the same geometrical analysis that
the type IIB backgrounds obtained through the S-Duality
were exact, provided the vector like
couplings in the metric were linear functions of the transverse 
coordinates \cite {ag}.
For the case when the couplings are arbitrary functions, 
there are explicit 
higher order contributions to the equations of motion 
\cite {tsetrev} and a straightforward
application of the geometrical method was not viable. 
For the K-model
backgrounds with purely NS-NS field content it was shown by Duval 
{\it et. al.} \cite {duval}
that the geometrical approach may be modified in the 
non-symmetric connection
framework of Osborn \cite {osborn}. For the one-loop 
beta functions, it was
possible to show in this fashion that the higher order
contributions from the metric
and the antisymmetric tensors to the equations of motion
cancel provided the vector like couplings
enter the theory in a chiral fashion. They further argued
that such cancellations will persist at
all higher orders in \ap provided 
certain field redefinitions exist in addition
such that all higher order terms may be expressed in a generic 
fashion.
This was later confirmed by an explicit higher order sigma model
analysis \cite {tsetrev}.

In the present article we apply the non-symmetric connection
framework to the type IIB K-model like superstring backgrounds 
obtained
by us in \cite {ag} . For these plane wave like type IIB backgrounds
there are no sigma model interpretations and the geometrical approach
alluded to earlier seems to be the only way of considering higher
order corrections.
A modification to the approach in \cite {duval}
is however required to accommodate the R-R sector.
Using this methodology we investigate the issue of
higher order contributions to the equations of motion 
for the case when
the vector like couplings are arbitrary functions. 
We observe that subject to the existence of 
certain field redefinitions and the
vector like couplings being chiral, the type 
IIB backgrounds \cite {ag} 
are also exact to all orders in \ap. 
We examine
the generic higher order contributions at two-loops explicitly .
From this we provide a strong 
argument for the existence of such field redefinitions.
We observe that for the backgrounds considered here, the $SL(2,R)$
transformations preserves the structure of \ap corrections subject
to the existence of appropriate field redefinitions.
 
The article is divided into five sections.
In Section-2 we briefly review the non symmetric connection approach
of Osborn and describe the application to K-model string backgrounds
due to Duval {\it et. al.} modified to include the dilaton field.
In Section-3 we present the solution generating technique and
describe the type IIB background obtained by us in \cite {ag}. 
In Section-4
we express the background field dynamics using a modified 
version of
the non symmetric connection which is consistent with the 
type IIB field
content. We also address the
issue of higher order corrections to the equations of 
motion through a 
purely geometrical analysis.
A summary and discussion of our results are presented 
in the concluding 
Section-5. \bigskip

\noindent {\bf 2. Non-Symmetric Connections and K-Model String 
Backgrounds.}
\bigskip

\noindent The non-symmetric connection approach 
\cite {duval,osborn} is a method
particularly suited for an unified description of the 
background field
dynamics of both the metric and the antisymmetric 
tensor field describing
a string background configuaration. We briefly
review this here and also include the dilaton contribution. 
The approach
proceeds as follows; if $G_{\mu\nu}, B_{\mu\nu}$ and $\phi$ describes
a consistent string background then one defines,
\be
{\cal E}_{\mu\nu}=G_{\mu\nu} + B_{\mu\nu}  \label {back}
\ee
and the non-symmetric connection is defined to be,
\be
\gam^{+\rho}_{~~ \mu\nu}=\gam^{\rho}_{~\mu\nu} 
+ H^{\rho}_{~\mu\nu} \label {non}
\ee
which may be expressed in the following way as;
\be
\gam^{+}_{\rho\mu\nu}=\hf\big[ \pa_{\nu}{\cal E}_{\rho\mu} + 
\pa_{\mu}{\cal E}_{\nu\rho}
- \pa_{\rho}{\cal E}_{\nu\mu}\big]. \label {gamp}
\ee
Here $H_{\mu\nu\rho}=\hf( \pa_{\mu}B_{\nu\rho} +\pa_{\nu}B_{\rho\mu}
+ \pa_{\rho}B_{\mu\nu})$ is the field strength corresponding to
the antisymmetric tensor $B_{\mu\nu}$.
The generalized curvature corresponding to these connections 
in eqn.(\ref {gamp})
are
\be
\rplus =\pa_{\mu}\gam^{+\rho}_{~\nu\sigma} - 
\pa_{\nu}\gam^{+\rho}_{~\mu\sigma} 
+\gam^{+\rho}_{~~\mu\alpha}
\gam^{+\alpha}_{~~\nu\sigma} - 
\gam^{+\rho}_{~~\nu\alpha}\gam^{+\alpha}_{~~\mu\sigma} 
\label {rplus}
.\ee
With this the generalized curvature $\rci$ is antisymmetric 
in the first and 
the second pair of indices.
The generalized curvature may also be expressed in terms 
of the usual 
Christoffel connections and the torsion in
the following way;
\be
\rplus = \rup 
+D_{\mu}H^{\rho}_{~ \nu\sigma} - D_{\nu}H^{\rho}_{~\mu\sigma}
+ H^{\rho}_{~\mu\alpha}H^{\alpha}_{~\nu\sigma}
-H^{\rho}_{~\nu\alpha}H^{\alpha}_{~\mu\sigma} \label {rplush}
.\ee
It is possible to derive a generalized Bianchi identity,
\be
D^{+}_{[\nu}R^{+\alpha}_{~~\mid \lambda\mid\beta\kappa]}
-2H^{\sigma}_{~[\nu\beta}
R^{+\alpha}_{~~\mid\lambda\mid\kappa]\sigma}=0. 
\label {bian} \ee
where $\mid\lambda\mid$ denotes no permutations on the index. 
It further follows
from $dH=0$ that,
\be
D^{+}_{\mu}H_{\nu\rho\sigma}={3\over2}R^{+}_{[\nu\rho\sigma]\mu}. 
\label {bian1}
\ee
The generalized Ricci tensor is then simply $R^{+}_{\mu\nu}=
R^{+\alpha}_{~~\mu\alpha\nu}$ and may be evaluated to be
\be
R^{+}_{\mu\nu}=R_{\mu\nu} + D_{\alpha}H^{\alpha}_{~\nu\mu}
-H^{\rho\alpha}_{~~\mu}H_{\nu\rho\alpha} \label {ricci}
.\ee
The symmetric part of (\ref {ricci}) then provides the usual metric 
one loop beta function
and the antisymmetric part is the corresponding expression for the
beta function of the antisymmetric tensor field, both without the usual 
dilaton part. So the one loop equations
of motion for the background after introducing the dilaton part
may be expressed as
\be
R^{+}_{\mu\nu} + D_{\mu}D_{\nu}\phi +D^{\lambda}\phi H_{\lambda\mu\nu}
=0. \label {gmu}
\ee

The K-model string backgrounds describing generalization of plane 
gravitational waves are given as follows \cite {tsetrev} 
with a Brinkmann
metric \cite {brink},
\be
ds^2=2dudv + 2A_{i}^{+}(u,x)dudx^i 
+ K(u,x) du^2 + dx^idx_i \label {kmet},
\ee
and an antisymmetric tensor field;
\be
B_{\mu \nu}=\left ( \begin{array}{ccc}
0 & 1 & A_{i}^{-}\\
-1 & 0 & 0\\
-A_{i}^{-} & 0 & 0\\
\end{array} \right ). \label {kb}
\ee
The vector like couplings $A^{\pm}_{i}$ plays a significant role 
in considering these models and provides gauge fields on 
compactification
to lower dimensions. Actualy they are expressed as
\be
A^{\pm}_{i}=A_{i}\pm {\bar A}_{i} \label {aipm}
\ee where 
$A_{i}$ and ${\bar A}_{i}$ are the couplings in the corresponding
bosonic sigma model \cite {tsetrev}. We also consider a dilaton
which is taken to be simply a function $\phi = \phi(u)$.

The $v$ isometry of the background is described by the Killing
vector $l^{\mu}$ which is given as $(0, 1, 0,.....0)$. 
The only non-zero connections computed from the metric (\ref {kmet}) 
are,
$\gam^{i}_{uu}$, $\gam^{v}_{uu}$, $\gam^{v}_{ui}$, 
$\gam^{j}_{ui}$ 
and
$\gam^{v}_{ij}$. Using these connections, it is easy to 
show that the
null killing vector $l^{\mu}$ is covariantly constant. 

The only non-zero independent components of the curvature
$\rc$ turn out to be \cite {ag} $R_{uiuj}$ and 
$R_{uijk}$. These may be evaluated to obtain;
\be
R_{uiuj}=\hf \pa_i\pa_j K - \hf \pa_{u}\big [ \pa_iA_j^+ + 
\pa_jA_i^+ \big ]         
-{1\over{4}}\delta^{mn}(F^{+}_{jm}F^{+}_{in}) \label {rui}
\ee
and
\be
R_{uijk}=\hf \pa_i \big [ F_{kj}^{+}\big ] \label{ruk}
\ee
where
\be
F_{jk}^{\pm}=\big( \pa_{j}A_{k}^{\pm} - \pa_{k} A_{j}^{\pm} \big).
\ee
The field strength of the antisymmetric tensor $B_{\mu\nu}$ 
in the standard
form is $H=\hf dB$ and this upon evaluation gives the only non zero 
independent component as,
\be
H_{uij}=-\hf F_{ij}^{-} \label {hu}.
\ee

It is now a straightforward exercise to cast this background in the 
non-symmetric connection framework described earlier. We essentialy
describe here the approach of Duval et. al. \cite {duval} and also
include the dilaton. Notice first, that the set of the non zero 
connections
remains the same as for the symmetric case for the backgrounds 
under consideration {\it cf} eqns.(\ref {kmet},\ref {kb}). 
The only connections
which are modified by the torsion contribution are $\gam^{v}_{ij}$
and $\gam^{j}_{ui}$. The non-zero components of the 
generalized curvature
are then
\be
R^{+i}_{~ jku}=-\hf\delta^{il}\pa_{k}\big[ \pa_{l}A^{+}_{j}
- \pa_{j}A^{+}_{l} - \pa_{l}A^{-}_{j} + \pa_{j}A^{-}_{l}\big]  
\label {rpjku}
\ee

\be
R^{+i}_{~ ujk}=\hf\delta^{il}\pa_{l}\big[ \pa_{k}A^{+}_{j}
- \pa_{j}A^{+}_{k} - \pa_{k}A^{-}_{j} + \pa_{j}A^{-}_{k}\big] 
\label {rpujk}
\ee
and
\bea
R^{+i}_{~ uju} &=& \delta^{il}\big[ \pa_{j}(\pa_{u}A_{l}^{+} 
+ \hf \pa_{l}K)
+ \hf \pa_{u}(\pa_{l}A^{+}_{j}
- \pa_{j}A^{+}_{l} - \pa_{j}A^{-}_{l} 
+ \pa_{l}A^{-}_{j})\big]\\ \nonumber
& + &{1\over 4} \big[ (F_{jk}^{+} + 
F_{jk}^{-}) (F_{k}^{+i} - F_{i}^{-k})\big] \label {rpuju}
\eea

The tree level equations of motion are then obtained 
from the generalized
Ricci tensors and the dilaton contributions. Using eqns. 
(\ref {kmet}, \ref {rpjku}, \ref {rpujk}, \ref {rpuju})
we obtain;
\be
R^{+}_{ju} 
=-\hf \pa^{i} \big[ \pa^{i}A_{j}^{+} - \pa_{j}A_{i}^{+} 
- \pa_{i}A_{j}^{-} + \pa_{j} A_{i}^{-} \big] =0 \label {ricju}
\ee
\be
R^{+}_{uj}=\hf \pa^{i} \big[ \pa^{i}A_{j}^{+} - \pa_{j}A_{i}^{+} 
+ \pa_{i}A_{j}^{-} - \pa_{j} A_{i}^{-} \big] =0 \label {ricuj}
\ee
and
\bea
R^{+}_{uu} + 2 \pa^{2}_{u}\phi &=& \pa^{i}\big( \pa_{u}A_{i}^{+} 
+\hf \pa_{i}K\big)
- {1\over4} \big( \pa^{k}A_{i}^{+} - 
\pa_{i}A_{k}^{+} \big)^{2}\\ \nonumber
&+& {1\over4} \big( \pa^{k}A_{i}^{-} - 
\pa_{i}A_{k}^{-} \big)^{2}\\ \nonumber
&=& 0 \label {ricuu},
\eea
where $2\pa_{u}^2\phi$ is the dilaton contribution. Notice that
the dilaton term coupling to the torsion in the beta function
equation, $D^{\lambda}\phi H_{\lambda\mu\nu}=0$ as a consequence
of
the Killing equations.

The arguments for the higher order corrections to these 
equations of motion
(\ref {ricju},\ref {ricuj}, \ref{ricuu}) as presented in 
\cite {duval}, proceeds as follows. 
These higher
order corrections are in fact all possible 
non-zero rank two tensors obtained
from appropriate contractions of the
generalized curvature, torsion, dilaton and their covariant
derivatives ( the covariant derivative of $H$ may be expressed 
in terms
of $R^+$ from eqn (\ref {bian}). This is of course because the 
eqns. of motion are field equations for rank two tensors.
If the vector like couplings 
satisfy the condition
\be
A^{+}_{i}=\pm A_{i}^{+} \label {condi}
\ee 
then it is obvious from eqn. (\ref {rpjku}, \ref {rpujk}) that
$R^+_{~uijk}$ vanishes. 
This translates to the condition, $G_{ui}=\pm B_{ui}$ for the metric
and the antisymmetric tensor components.
From the bosonic sigma model point of view this essentialy means
that the couplings $A_{i}=0$ and ${\bar A}_{i}=0$ 
from eqn. (\ref{aipm}) i.e.
the vector like couplings in the sigma model are chiral.
With this condition ( for eg. we will consider the positive sign )
the conformal invariance conditions at one loop reduce 
to only two equations
namely;
\be
R^{+}_{uj}=\pa^{i}\big( \pa_{i}A_{j}^{+} - 
\pa_{j} A^{+}_{i} \big )=0 \label {chiric}
\ee
and
\be 
R^{+}_{uu} + \pa^{2}_{u}\phi= \pa^{i} \big( \pa_{u}A_{i}^{+} 
- \hf \pa_{i}K\big )
+ \pa^{2}_{u}\phi =0.
\ee
In consequence of the vanishing of $R^+_{uijk}$, the generalized 
curvature with three 
spatial indices, all terms involving $R^+$ and $H$, of 
rank two are identicaly
zero. This is
because they require contraction of at least one $u$ index which is
not allowed by the specific form of the metric. Next in the 
terms with $R^+$
and its derivatives, notice that $D_{v}^+$ operation on the 
background is
zero valued, as both $\gam_{v.}^{+.}$ and $H_{v..}$ 
( where the ellipses denote other indices)
are zero owing to the isometry of the background in $v$.
In consequence the operation $D^{+u}$ is
also zero from the form of the metric. Hence all such terms 
involving these derivatives are identicaly 
zero as they require contractions of the index $u$.
Simmilar considerations show that terms involving 
$D^+R^+$ and $D^+H^+$
are also vanishing.

Finally we consider terms in single $R^{+}$ of the form
$D^{+\mu}D^{+\rho}R^{+}_{\mu\nu\rho\sigma}$. 
This may be expressed in terms of the derivatives of the
generalized Ricci tensor using the generalized Bianchi identity. 
It is possible to show using this 
\cite {duval} that all such terms vanish. 
Similarly using the Killing equations for the other backgrounds
it may be shown that
all higher order terms involving the dilaton and its 
derivatives are also 
identicaly zero \cite {horo,duval,asg,ag}. 
It turns out, at least at two loop level that
all possible non-zero higher order contributions may be 
reduced to the
form \cite {tsetrev}
\be
\beta_{\mu\nu}= Y^{+\lambda\rho\sigma}_{~~~~~\nu}
R^{+}_{~\mu\sigma\lambda\rho} \label {yr}
\ee
where
$Y^{+\lambda\rho\sigma}_{~~~~~\nu}$ is a function of $R^{+}$ and
derivatives of $H$. If appropriate field redefinitions 
exist such that
this is continued at higher orders and $R_{uijk}^{+}$ is zero
from eqn. (\ref {condi}) then
all higher order terms vanish. This is because at least one of 
the indices
to be contracted in (\ref {yr})
must be $u$ and it has been shown that all such contractions are
identicaly zero.
So we may conclude that the K-model string backgrounds are exact
subject to these conditions. This conclusion has been corroborated
by an explicit sigma model based proof \cite {tsetrev}. 
This cancellation is of course a result of the covariantly constant
null Killing isometry which these backgrounds admit and the
condition (\ref {condi}).
This concludes the description
of K-model string background in the non-symmetric connection
framework \cite {duval}.
In the next section we describe the technique of generating 
non trivial
R-R sector through the S-duality transformations and cast the 
type IIb
backgrounds obtained in the non-symetric framework prior 
to examining
the issue of all higher order corrections to the equations 
of motion.
\bigskip

\noindent{\bf 3. Type IIB K-model Superstring Backgrounds.} 

\noindent The complete massless bosonic field content of 
type IIb superstring
backgrounds consists of the string frame metric $G_{\mu\nu}$, two
2-form gauge fields $B_{\mu\nu}^{(A)}$ with $A=(1, 2)$, two scalars
$\phi$ and $\chi$ from the NS-NS and the R-R sectors respectively
and a real self-dual 4-form $D_{\mu\nu\rho\sigma}$. 
As mentioned earlier
the type IIB superstrings in D=10 have a global $SL(2,R)$ symmetry
of the equations of motion of the effective supergravity 
theory \cite {hull}.
The action of the $SL(2,R)$ transformations on the background fields
may be described as follows.
We define the complex scalar field $\lambda=\chi + ie^{\phi}$. 
Then if
$\Lambda$ is an $SL(2,R)$ matrix such that
\be
\Lambda = \left( \begin{array}{cc}
d & c\\
b & a\\
\end{array}\right) , \label{s4}
\ee
with $ad-bc=1$, the action on the type IIb background fields are
\be
G_{\mu \nu}^{\prime}=\mid c\lambda + d \mid G_{\mu \nu}, \label{s1}
\ee

\be 
\lambda^{\prime}= {{a \lambda + b}\over {c \lambda + d}}, \label{s2}
\ee
and 
\be
\hi^{\prime (A)}=\Lambda \hi^{(A)}, \label{s3}
\ee
where $H^{(A)}$ are the field strengths corresponding 
to the two 2-forms
$B^{(A)}$, (A=1, 2), from the NS-NS and the R-R sectors 
respectively.

We may consider the K-model backgrounds described 
in the last section to be
a special case of the type IIB with the R-R fields $B^{(2)}=0$, 
$\chi=0$ and $D=0$. 
So we have $\lambda=ie^{\phi}$ as $\chi=0$ and implementing 
the $SL(2,R)$ 
transformations (\ref {s1}, \ref {s2}, \ref {s3}) on this background 
we obtain
the complete type IIB background with the following field content;
\be 
G_{\mu \nu}^{\prime}(u, x)=f(u) G_{\mu \nu}(u, x), \label{gp}
\ee
where $f(u)={\big [d^{2} + c^{2}e^{-2\phi (u)}\big ]}^{\hf}$
and
\be 
\chi^{\prime}(u)=
{1\over {{f(u)}^{2}}}\big [  db + ac\ e^{-2\phi}\big], \label{xp}
\ee
\be
\phi^{\prime}(u)=\phi (u) + 2\ ln\ f(u). \label{pp}
\ee
For
the 3-form field strength $H^{(A)}$, $(A=1, 2)$ we have
\be
\hi^{\prime (1)}=d \hi^{(1)}, \label{hp1}
\ee
and 
\be
\hi^{\prime (2)}=b \hi^{(1)}. \label{hp2}
\ee
After a rescaling $f(u)du=dU$ and rewriting $U$ as $u$
leads to the general form; 
\be
ds^{2}=2 dudv + 2 f(u)dx^{i}dx_{i} + 2 A_i^+(u,x)dudx^i
+ K (u,x) du^{2}. \label{ndsf}
\ee
In subsequent discussions we drop the primes on the type IIB
background fields generated by the $SL(2,R)$ transformations 
from the
K-model backgrounds and also employ the same notation for the
transformed functions $A^{\pm}_{i}$ and $K$.

Notice that the null Killing isometry has been preserved 
by the S-duality.
Using the expressions for the metric from eqn. (\ref {ndsf}) 
we
observe that the only non-zero components of the Christoffel 
connections are, 
$\gam^{v}_{uu}$, $\gam^{i}_{uu}$, $\gam^{v}_{ui}$,
$\gam^{j}_{ui}$ and $\gam^{v}_{ij}$. 
From the non-zero connections it follows that the
null Killing isometry is still covariantly constant which is expected
as the S-duality respects space-time geometries.
The only non-zero independent components of the curvature tensor
\cite {ag} turn out to be as follows,
\bea
R_{uiuj}&=& \hf \pa_{i}\pa_{j}K - \hf \pa_{u}
\big[ \pa_{i}A^{+}_{j} +
\pa_{j}A_{i}^{+}\big] \\ \nonumber
&+& \hf \pa_u^2f\delta_{ij} -{1\over{4f^2}} 
\big[ \delta_{ij}(\pa_{u}f)^{2}
+ F_{jm}^{+}F_{i}^{+m}\big] \label {rsij}
\eea
and
\be
R_{uijk}=\hf \pa_i \big [ \pa_kA_j^+ - \pa_jA_k^+\big ]. 
\label {rsijk} 
\ee
Notice that expression for $R_{uijk}$ is unchanged from that 
for the
purely NS-NS background. Having obtained the complete 
type IIB background 
with non-zero R-R field content, in the next Section we proceed 
to cast these into the
non-symmetric connection framework described earlier.
\bigskip

\noindent {\bf 4. Type IIB Background and the Non-Symmetric 
Connection Approach.}

\noindent In this section we present the non-symmetric 
connection approach
for the type IIB backgrounds obtained by us \cite {ag}
with non-zero R-R field content. The definition of the 
non-symmetric connection in eqn. (\ref {non}) requires to be 
modified due to the the two torsions arising from the NS-NS and the
R-R sectors. As the torsions are additive the simplest modification
which suggests itself is,
\be
{\cal H}_{\lambda\mu\nu}= H^{(1)}_{\lambda\mu\nu} +
 H^{(2)}_{\lambda\mu\nu} \label {scale}
\ee
$H^1=dH$ and $H^2=bH$ where $d$ and $b$ are just the elements 
of the
$SL(2,R)$ matrix $\Lambda$. Notice that such a definition is
consistent with general covariance and space-time tensor gauge
invariance. This also preserves all the relations for the
non-symmetric connection presented in Section-2 including the
generalized Bianchi identities
with ${\cal H}$ replacing $H$ in (\ref {non}). 
In fact one may use a linear
combination of the two torsions also where the coefficients
would be restricted by the generalized Bianchi identity. Notice
that for the backgrounds being analysed, the definition 
(\ref {scale})
corresponds to a simple constant scaling by the quantity $(d+b)$
of the starting NS-NS
2-form and consequently implies a simmilar scaling on the 
antisymmetric tensor components.

We now proceed to explicitly compute the generalized curvatures
and Ricci tensors for the background described in the last section. 
As earlier the set of non-zero connections remain the same
as in the symmetric case and the modifications to 
their forms identical.
The only non-zero independent components of the generalized 
curvature are 
\be
R^{+i}_{jku}=-{{1}\over{2f(u)}} 
\delta^{il}\pa_{k}\big[ \pa_{l}A^{+}_{j}
- \pa_{j}A^{+}_{l} - \pa_{l}A^{-}_{j} 
+ \pa_{j}A^{-}_{l}\big] \label {rspjku}
\ee
\be
R^{+i}_{ujk}={{1}\over{2f(u)}} 
\delta^{il}\pa_{l}\big[ \pa_{k}A^{+}_{j}
- \pa_{j}A^{+}_{k} + \pa_{k}A^{-}_{j} 
- \pa_{j}A^{-}_{k}\big] \label {rspujk}
\ee
and
\bea 
R^{+i}_{~uju} &=& {\delta^{il}\over 2f}\big[  \pa_{l}\pa_{j}K - 
\pa_{u}(\pa_{l}A_{j}^{+}
+\pa_{j}A^{+}_{l}) + \pa_{u}(\pa_{l}A_{j}^{-} 
- \pa_{j}A^{-}_{l}) \\ \nonumber
&+&\pa_{u}^2f\delta_{jl} -\hf \delta_{jl} (\pa_{u}f)^2 \\ \nonumber
&+&{1\over2f}(F^{+n}_{j} 
+ F^{-n}_{j})(F^{+}_{ln} - F^{-}_{ln})\big] \label {rspuju}
\eea 
Where now the $A_{j}^{-}$ have been suitably scaled to 
accomodate for the
combination defined in eqn. (\ref {scale}). 
The tree level supergravity equations of motion 
for the metric and the antisymmetric tensor may now be 
once again expressed in terms of the generalized
Ricci tensors combined with the appropriate scalar field
contributions. The equations of motion for the scalar fields must be 
added on by hand separately.
The torsion equations would be identical in form to
eqns. (\ref {ricju}, \ref {ricuj}), except that the $A_{j}^{-}$ have
now been suitably scaled due to relation (\ref {scale}).
The only change will involve the metric
equation where additional contribution from the R-R scalar must be
included. The $(uu)$ component of the Ricci tensor is just
\be
R^{+}_{uu}={1\over{f^{2}}}\big[ {1\over2}\pa^{i}\pa_{i}K +4 \pa_{u}^2 
f -2 (\pa_{u}f)^2
+{1\over4}( F^{+}_{nl}-F^{-}_{nl})^2\big] \label {ruus}
\ee
These along with appropriate scalar contributions
essentialy describes the tree level type IIB supergravity 
equations of motion for the backgrounds under consideration.

Notice that the tree level equations of type IIB 
supergravity in the string
frame metric in ref. \cite {berg} reduces for the backgrounds under 
consideration to just four independent equations as $F_{5}=0$,
the ones for the torsion
being identical. These are the equations for the metric, 
the two scalars
and the antisymmetric tensor
In particular we have the equation for the metric as
\be
R_{uu}= 4\pa_{u}\phi\pa_{u}\phi 
- {{e^{2\phi}}\over 2} \pa_{u}\chi\pa_{u}\chi
+ w(u) H^2_{uu} =0 \label {2beqn}
\ee
where $w(u)$ is a complex scalar function of $u$. It is possible
to obtain (\ref {2beqn}) from (\ref {ruus}) with a suitable scaling
of the metric and addition of appropriate scalar contributions.
Notice that the contribution $D^{\rho}H_{\rho\mu\nu}$
to the equation is 
zero owing to the torsion equations of motion.
For $\phi\ \ \chi$ constant we have from ref. \cite {berg} that
the function $w(u)$ is a constant and with appropriate scaling
we get back eqn. (\ref {ruus}).
The other equations from ref. \cite {berg} for the background under
consideration are;
\be
D^2\phi=0.
\ee
\be
D^2\chi=0.
\ee
\be
D^{\mu}H_{\mu\nu\rho}=0.
\ee
and
\be
F_{5}(D)={\tilde F}_{5}(D)=0
\ee

Having presented the tree level equations of motion for the 
effective type
IIB supergravity theory
we now proceed to address the question of the higher order \ap
contributions to these
equations. All the arguments presented in Section-2 for the
higher order corrections
to the purely NS-NS case are also valid
for the type IIB case.
This is owing to the fact that the 2-forms in the NS-NS and 
the R-R sectors
turn out to be just simple scalar multiples of the 
starting NS-NS 2-form
and also because the generalized curvature has the same 
non zero components.
Using the Killing equations it may be shown that all 
possible scalar
and rank five antisymmetric tensor corrections, if any to 
the trivialy self
dual condition for $F_5$ are also zero \cite {ag}.
If we now adopt the condition for the 
chirality of the vector like couplings, namely
\be
A_{i}^{+}=\pm A_{i}^{-}
,\ee
or equivalently $G_{ui}=B_{ui}$ we have 
the curvature component $ R^{+}_{~ uijk}$ with three transverse
indices identicaly zero. 
Using the arguments outlined in Section-2 for the purely NS-NS
backgrounds it may be shown that the type IIB backgrounds 
under consideration are also exact to all orders in \ap.
Obviously this is subject to the fact that all higher order
terms as earlier, may be expressed in the form 
$Y^{+\lambda\rho\sigma}_{~~~~~\nu}R^{+}_{~ \mu\sigma\lambda\rho}$
through
field redefinitions.

The argument is not completely rigourous in the absence 
of an explicit 
sigma model description for the non-p-brane like
type IIB superstring backgrounds which are considered here.
However, a stronger argument for the existence of the 
field redfinitions
may be obtained by explicitly checking the generic non-zero 
higher order correction terms that may arise. 
Notice that the only higher order terms that may occur are 
in the equation
following from the $(uu)$ component of the Ricci tensor owing 
to the form 
of the background metric. Following arguments simmilar to those
in Section-2 it may be shown that all higher order corrections
to the other equations of motion vanish identicaly.

The only non-zero higher order correction at two-loop order to
eqn. (\ref {2beqn})
that may arise are from $R_{uijk}R_{u}^{~ ijk}$ and 
$D_{k}H^{(A)}_{uij}
D^{k}H_{u}^{(B)~ ij}$ where $(A, B= 1, 2)$ refers to the NS-NS and the
R-R sector and $R$ is the usual curvature. This is obvious from the
form of the background as all other higher order terms vanish as
a consequence of the covariantly constant null Killing isometry.
Assuming that the coefficients are just numerical
constants we have the explicit higher order term;
\be
\beta^{2}_{uu}= pR_{uijk}R_{u}^{~ ijk} +  q_{AB}D_{k}H^{(A)}_{uij}
D^{k}H_{u}^{(B)~ ij} 
\ee
where $p$ and $q_{AB}$ are constants.
This reduces for the background in our case to
\be
\beta^{2}_{uu}=pR_{uijk}R_{u}^{~ ijk} +  qD_{k}H_{uij}
D^{k}H_{u}^{~ ij} \label {2loop}
\ee
where $H$ is the starting NS-NS
torsion and $q$ is another constant. 
The terms in eqn. (\ref {2loop}) are the only 
non-zero terms in the contraction,
$R_{uijk}^+ R_{u}^{+ ijk}$.
This strongly suggests that under suitable redefinitions eqn.
(\ref {2loop}) reduces to $R_{uijk}^+ R_{u}^{+ ijk}$.
However
as $R^+_{uijk}=0$ with the condition defined by eqn. (\ref
{condi}) the contraction of the two generalized curvatures vanish.
This shows that all higher order contributions at two-loop
orders to the equations of motion are identicaly zero
subject to appropriate field redefinitions.
It is quite plausible that this
generalizes to arbitrary higher orders
for the background under consideration. This is because, 
firstly the corrections appear only for the $uu$ component 
of the metric
equation and all higher order terms must arise from 
terms of the schematic
form $D...R_{uijk}D...R_{uijk}$ and
$D...H_{uij}D...H_{uij}$ suitably contracted.
These may always be combined under suitable
field redefinition to give terms of the generic 
form as defined earlier.
The results indicate that the type IIB S-duality 
transformations preserve
the \ap structure of the backgrounds considered here. Whether this
assertion is true in general is an interesting issue to explore.
\bigskip

\noindent {\bf 5. Summary and Discussions.}

\noindent To summarize, we have examined the issue of 
higher order corrections
to the equations of motion of type IIB superstring 
backgrounds obtained
by us recently \cite {ag}. 
These ten dimensional backgrounds were obtained through the
S-duality transformations of plane wave K-model backgrounds 
with purely 
NS-NS field content.
They describe the most general class of type IIB 
superstring backgrounds with a covariantly constant null 
Killing isometry
and a flat transverse part in ten dimensions involving non-zero
Ramond-Ramond field content.
A geometrical approach in the non-symmetric
connection framework was adopted for this exercise
as a sigma model interpretation
for these backgrounds are still obscure.
The non-symmetric connection approach was generalized to these
type IIB superstring backgrounds by using a additive combination
of the two torsions from the NS-NS and the R-R sectors.
From a geometrical analysis of the tensor structure of the
possible higher order terms we show that the type IIB K-model
superstring backgrounds are exact to all orders in \ap.
This is subject to the condition that the vector like K-model
couplings enter the theory in a chiral fashion and the existence
of certain field redefinitions.
We obtain a strong argument in favour of the existence of the field 
redefinitions through an explicit representation of 
the generic higher order
corrections at two-loops. With the chirality condition 
the backgrounds
obtained by us are the type IIB analogs of chiral plane waves
in ref. \cite {tsetrev}.

Our results indicate that the $SL(2,R)$ S-duality of type IIB
in ten dimensions preserves the structure of \ap corrections
of the chiral plane wave backgrounds. Of course this is a direct
consequence of the covariantly constant null Killing isometry
and interelation between the metric and the antisymmetric-tensor
for the special backgrounds that we have considered.
Whether such an assertion is true in general seems to be an
interesting issue to investigate.
A general proof seems quite non-trivial
especialy as the structures of higher order corrections
are not known explicitly. 
These are quite difficult to obtain even for
p-brane backgrounds for which an explicit Dirichlet sigma model
description is available \cite {pol}.
Furthermore a sigma model
interpretation for the non-brane like backgrounds with R-R fields
discussed here is still an obscure issue which needs to be addressed.
Our results strongly indicate that such an exact conformal field
theory description should exist. Some of these type IIB backgrounds
with non-trivial R-R fields have been recently 
identified with Toda models
in the context of their applications to string 
cosmology \cite {ovrut}

It has been shown \cite {berg1} that the NS-NS chiral
plane wave backgrounds may be embedded both in heterotic and 
type I
superstrings and for the latter case preserves only half 
of the space-time
supersymmetry in ten dimensions. 
The status of unbroken space-time 
supersymmetry would be also interesting to investigate 
in the full type IIB
case discussed here. The results will have implications for 
these type IIB backgrounds
to be solutions in presence of local string loop corrections
as well. These issues are currently under consideration.
\bigskip

\noindent {\bf Acknowledgements:} I would like to thank Alok Kumar
for many discussions and collaborations in refs. 
\cite {asg,ag}. I would also like to thank
Parthasarathi Majumdar, T. Jayaraman
for discussions and
suggestions and Stefan Forste for
useful correspondence.
Furthermore I would like to acknowledge the warm hospitality at
Center for Theoretical Studies, Bangalore where part of this
work was completed.

\vfil\eject

\end{document}